\documentclass[a4paper,11pt]{article}
\usepackage{pos}
\usepackage{caption}
\usepackage{subcaption}

\newcommand{\nn}{\nonumber}
\newcommand{\mhh}{m_{hh}}

\newcommand{\ct}{c_t}
\newcommand{\ctt}{c_{tt}}
\newcommand{\chhh}{c_{hhh}}
\newcommand{\cg}{c_{ggh}}
\newcommand{\cgg}{c_{gghh}}

\title{Two Higgs bosons, two loops, x+2 operators}

\author*[a]{Gudrun Heinrich}
\author[a]{Jannis Lang}
\author[b]{Ludovic Scyboz}

\affiliation[a]{Institute for Theoretical Physics, Karlsruhe Institute of Technology (KIT),\\
  76131 Karlsruhe, Germany}
\affiliation[b]{School of Physics and Astronomy, Monash University,
Wellington Rd, Clayton VIC-3800, Australia}

\emailAdd{gudrun.heinrich@kit.edu}
\emailAdd{jannis.lang@kit.edu}
\emailAdd{ludovic.scyboz@monash.edu}

\abstract{We discuss the combination of NLO QCD corrections with
  operators of canonical dimension six within Standard Model Effective Field
  Theory (SMEFT), as well as within non-linear Effective Field
  Theory (HEFT) for Higgs-boson pair production in gluon
  fusion. Particular emphasis will be put on the identification of leading and subleading operators contributing to this process.}

\FullConference{16th International Symposium on Radiative Corrections: Applications of Quantum Field Theory to Phenomenology (
RADCOR2023)\\
28th May - 2nd June, 2023\\
Crieff, Scotland, UK\\}

\begin{document}
\maketitle

\section{Motivation}

Since the spectacular discovery of the Higgs boson in 2012, physics
beyond the Standard Model~(SM) did not manifest itself by the
observation of new particles at energies that are directly accessible at the Large Hadron Collider (LHC).
However, indirect signs of new physics at larger energy scales can also be compelling if the uncertainties on both measurements and theory predictions are well under control.

If there is an energy gap between a new physics scale $\Lambda$ and the electroweak scale, Effective Field Theory (EFT) is the appropriate framework to describe effective interactions between the known fields, resulting from heavier degrees of freedom that have been integrated out.
However, for high precision predictions, perturbative higher order corrections need to be taken into account, and it is a non-trivial task to combine QCD and electroweak (EW) corrections with EFT parametrisations of new physics.
The work reported here addresses this subject in the context of Higgs boson pair production in gluon fusion, which is the prime process to constrain the trilinear Higgs boson self-coupling.
In the SM, the leading order for this process already proceeds via heavy fermion loops. Therefore, the NLO corrections involve two-loop four-point integrals with at least two mass scales ($m_h, m_t$) and thus are challenging to calculate.

The NLO QCD corrections in the SM are
available~\cite{Borowka:2016ehy,Borowka:2016ypz,Baglio:2018lrj,Baglio:2020ini,Davies:2019dfy,Bagnaschi:2023rbx},
corrections beyond NLO, using large-$m_t$ expansions  or the high energy limit, also have been
calculated~\cite{AH:2022elh,Chen:2019fhs,Chen:2019lzz,Davies:2021kex,Grazzini:2018bsd,Deflorian:2018eng,deFlorian:2016uhr,Grigo:2015dia}.
Partial three-loop results~\cite{Davies:2023obx} as well as
EW corrections also have emerged recently~\cite{Davies:2023npk,Davies:2022ram,Muhlleitner:2022ijf}.
The results of \cite{Borowka:2016ehy} have been implemented in into the {\tt Powheg-Box-V2}~\cite{Nason:2004rx,Frixione:2007vw,Alioli:2010xd} event generator, first for the SM only~\cite{Heinrich:2017kxx}, then also for $\kappa_\lambda$ variations~\cite{Heinrich:2019bkc} as well as for the leading operators contributing to this process in HEFT~\cite{Buchalla:2018yce,Heinrich:2020ckp} and SMEFT~\cite{Heinrich:2022idm}.
In Ref.~\cite{deFlorian:2017qfk} the combination of NNLO corrections in an $m_t$-improved heavy top limit (HTL)
has been performed including anomalous couplings, extending earlier work at NLO in the $m_t$-improved HTL~\cite{Grober:2015cwa,Grober:2017gut}.
The work of~\cite{deFlorian:2017qfk}  has been combined with the full NLO corrections within non-linear EFT of
Ref.~\cite{Buchalla:2018yce}  to provide approximate NNLO predictions
in Ref.~\cite{deFlorian:2021azd}, dubbed NNLO$^\prime$.

However, when combining higher order corrections with EFT expansions, new sources of uncertainties arise.
Already the choice of the EFT counting scheme that is used to assess which operators are the leading ones implicitly makes some (minimal) UV assumptions; the truncation order of the SMEFT expansion in $1/\Lambda$ or the (implicit or explicit) scale choice in the renormalisation group running of the Wilson coefficients also introduce additional sources of uncertainties.
Some of these uncertainties will be discussed in the following in the context of Higgs boson pair production in gluon fusion.

\section{Effective Field Theory descriptions of Higgs boson pair production in gluon fusion}

\subsection{SMEFT}

In Standard Model Effective Field Theory
(SMEFT)~\cite{Buchmuller:1985jz,Grzadkowski:2010es,Brivio:2017vri,Manohar:2018aog,Isidori:2023pyp}
it is assumed that the physical Higgs boson is part of a doublet
transforming linearly under $SU(2)_L\times U(1)$.
The effects of interactions at a new physics scale $\Lambda$ are
parametrised as an expansion in inverse powers of $\Lambda$, with operators
$\mathcal{O}_i$ of canonical dimension larger than four and corresponding Wilson
coefficients $C_i$,
\begin{equation}
\mathcal{L}_\text{SMEFT} = \mathcal{L}_\text{SM} + \sum_{i}\frac{C_i^{(6)}}{\Lambda^2}\mathcal{O}_i^{\rm{dim6}} +{\cal O}(\frac{1}{\Lambda^3})\; .
\label{eq:Ldim6}
\end{equation}
As an operator basis up to canonical dimension six,
the so-called Warsaw basis~\cite{Grzadkowski:2010es} is commonly used, where the
terms relevant to Higgs boson pair production in gluon fusion are given by
\begin{equation}
\begin{split}
  \Delta\mathcal{L}_{\text{Warsaw}}&=\frac{C_{H,\Box}}{\Lambda^2} (\phi^{\dagger} \phi)\Box (\phi^{\dagger } \phi)+ \frac{C_{H D}}{\Lambda^2}(\phi^{\dagger} D_{\mu}\phi)^*(\phi^{\dagger}D^{\mu}\phi)+ \frac{C_H}{\Lambda^2} (\phi^{\dagger}\phi)^3 \\ &+\left( \frac{C_{uH}}{\Lambda^2} \phi^{\dagger}{\phi}\bar{q}_L\phi^c t_R + h.c.\right)+\frac{C_{H G}}{\Lambda^2} \phi^{\dagger} \phi G_{\mu\nu}^a G^{\mu\nu,a}\\
  &+\left(\frac{C_{t G}}{\Lambda^2} \,\bar q_L\sigma^{\mu\nu}T^aG_{\mu\nu}^a\tilde\phi t_R +{\rm h.c.}\right)\;.  \label{eq:warsaw}
\end{split}
\end{equation}

If we assume a renormalisable, weakly coupled UV theory, 
the dipole operator
\begin{align}
  {\cal O}_{tG}=\bar q_L\sigma^{\mu\nu}T^aG_{\mu\nu}^a\tilde\phi t_R +{\rm h.c.}
\end{align} and the operator
\begin{align}
  {\cal O}_{HG}=\phi^{\dagger} \phi G_{\mu\nu}^a G^{\mu\nu,a}
\end{align}
in eq.~(\ref{eq:warsaw}) are loop-generated and therefore their coefficients are expected to be
suppressed by a  factor $1/(4\pi)^2$ relative to the other contributions, as will be explained in Section~\ref{sec:counting}.

Expanding the Higgs doublet in eq.~\eqref{eq:warsaw} around its vacuum
expectation value and applying a field redefinition to the physical Higgs boson
\begin{align}
h\to h+v^2\frac{C_{H,\textrm{kin}}}{\Lambda^2}\left(h+\frac{h^2}{v}+\frac{h^3}{3v^2}\right)\;,\label{eq:field_redefinition}
\end{align}
with $$C_{H,\textrm{kin}}:=C_{H,\Box}-\frac{1}{4}\,C_{HD}\;,$$
the Higgs kinetic term acquires its canonical form (up to ${\cal O}\left(\Lambda^{-4}\right)$ terms).

\subsection{HEFT}

Higgs Effective Field Theory
(HEFT)~\cite{Feruglio:1992wf,Burgess:1999ha,Grinstein:2007iv,Contino:2010mh,Alonso:2012px,Buchalla:2013rka},
also called non-linear Effective Field Theory or Electroweak Chiral Lagrangian (EWChL), 
 is based on an expansion in terms of loop orders $L$, which also can be
 formulated by counting the chiral dimension $d_\chi=2L+2$~\cite{Weinberg:1978kz,Buchalla:2013eza,Krause:2016uhw}.
 The expansion parameter in HEFT is given by $f^2/\Lambda^2\simeq \frac{1}{16\pi^2}$, where $f$ is a typical energy scale
at which the EFT expansion is valid, such as the pion decay constant in chiral
perturbation theory,
\begin{align}
  {\cal L}_{\chi}={\cal L}_{(d_\chi=2)}+\sum_{L=1}^\infty\sum_i\left(\frac{1}{16\pi^2}\right)^L c_i^{(L)} O^{(L)}_i\;.
  \label{eq:loop_expansion}
  \end{align}

The HEFT Lagrangian relevant to Higgs-boson pair production in gluon fusion up to $d_\chi=4$ can
be parametrised by five uncorrelated anomalous couplings $c_i$ as
follows~\cite{Buchalla:2018yce}
\begin{align}
\Delta{\cal L}_{\text{HEFT}}=
-m_t\left(c_t\frac{h}{v}+c_{tt}\frac{h^2}{v^2}\right)\,\bar{t}\,t -
c_{hhh} \frac{m_h^2}{2v} h^3+\frac{\alpha_s}{8\pi} \left( c_{ggh} \frac{h}{v}+
c_{gghh}\frac{h^2}{v^2}  \right)\, G^a_{\mu \nu} G^{a,\mu \nu}\;.
\label{eq:ewchl}
\end{align}

In the broken phase,  the anomalous couplings in HEFT and SMEFT can be related through a comparison of the coefficients of the corresponding terms in the Lagrangian, which leads to the expressions given in Table~\ref{tab:translation}.
However, such a ``translation'' should be used with great care, as the two EFTs rely on different assumptions and
operator counting schemes.

\begin{table}[htb]
\begin{center}
\begin{tabular}{ |c |c| }
\hline
HEFT&Warsaw\\
\hline
$c_{hhh}$ & $1-2\frac{v^2}{\Lambda^2}\frac{v^2}{m_h^2}\,C_H+3\frac{v^2}{\Lambda^2}\,C_{H,\textrm{kin}}$ \\
\hline
$c_t$ &  $1+\frac{v^2}{\Lambda^2}\,C_{H,\textrm{kin}} - \frac{v^2}{\Lambda^2} \frac{v}{\sqrt{2} m_t}\,C_{uH}$\\
\hline
$ c_{tt} $ & $-\frac{v^2}{\Lambda^2} \frac{3 v}{2\sqrt{2} m_t}\,C_{uH} + \frac{v^2}{\Lambda^2}\,C_{H,\textrm{kin}}$\\
\hline
$c_{ggh}$ &  $\frac{v^2}{\Lambda^2} \frac{8\pi }{\alpha_s} \,C_{HG}$ \\
\hline
$c_{gghh}$ &  $\frac{v^2}{\Lambda^2}\frac{4\pi}{\alpha_s} \,C_{HG}$ \\
\hline
\end{tabular}
\end{center}
\caption{Translation at Lagrangian level between different operator basis choices.\label{tab:translation}}
\end{table}

Note that in the Warsaw basis $C_{HG}$
implicitly contains a factor of $\alpha_s(\mu)$ relative to $\cg$ and $\cgg$ and
therefore the translation becomes scale-dependent.

\section{Operator counting}
\label{sec:counting}

A bottom-up EFT of unknown physics at so far unreachable energy scales has to be built on the following concepts:
the relevant local and global symmetries, the low-energy particle content and
a systematic power counting.
The latter relies on the presence of one or several expansion parameters which need to be small for the series to make sense, therefore a mass gap between the unknown dynamics residing at a large mass scale $\Lambda$ and the electroweak scale is mandatory.
Assumptions concerning the question of weakly versus strongly coupled underlying dynamics also influence the power counting scheme.
In the SMEFT, the dimensionless expansion parameter is given by $E^2/\Lambda^2$.
However, quantum corrections are inherently linked to loop counting,
therefore considerations about loop-induced versus potentially tree-level induced operators are relevant.
The systematics of power counting taking into account loop factors recently has been addressed in detail in Ref.~\cite{Buchalla:2022vjp}, building on earlier work~\cite{Arzt:1994gp,Einhorn:2013kja}.

We can organise the loop counting by defining an intermediate scale $f=\Lambda/4\pi$, reminiscent of the pion decay constant $f_\pi$ in chiral perturbation theory for pions as condensates of strongly coupled fermions.
Therefore, in strongly coupled theories, $f$ has a physical interpretation, while in weakly coupled theories such as SMEFT, it is rather a book-keeping parameter. For energies $E\sim f$ we then have $E^2/\Lambda^2\approx f^2/\Lambda^2=1/16\pi^2$, corresponding to a loop factor.
A general operator ${\cal O}$ in the EFT Lagrangian is composed of scalar fields $\phi$, vector fields $A_\mu$, fermion fields $\psi$, derivatives and generic weak couplings $\kappa$, each raised to some power:
\begin{align}
C\cdot {\cal O}\sim C\cdot \partial^{N_p}\kappa^{N_\kappa}\phi^{N_\phi}A^{N_A}\psi^{N_\psi}\;,
\end{align}
where the  EFT power counting scheme guides the estimation of the size of the coefficient $C$.
Assuming that the theory above the scale $\Lambda$ is renormalisable, some operators can only be loop-generated, therefore
the size in general will depend on both, the canonical dimension $d_c$ and the loop order $L$, which can  be expressed in terms of the chiral dimension $d_\chi$. More explicitly, one finds~\cite{Buchalla:2022vjp}
\begin{align}
  C(d_c, d_\chi)&=\frac{f^{4-d_c}}{(4\pi)^{d\chi-2}}=\frac{1}{(4\pi)^{d_\chi-d_c-2}}\frac{1}{\Lambda^{d_c-4}}\;,\label{eq:Wcoeff}\\
  d_c&=N_p+\frac{3}{2}N_\psi+N_\phi+N_A\;,\;
  d_\chi=N_p+\frac{1}{2}N_\psi+N_\kappa=2L+2\;.\nn
\end{align} 
Specifying to canonical dimension six, we find
\begin{align}
C(d_c=6, d_\chi)=\frac{1}{\Lambda^2}\left( \frac{1}{16\pi^2}\right)^\frac{d_\chi-4}{2}\;.
\end{align}
Therefore, operators with $d_\chi= 6$ will come with a loop suppression factor $1/(4\pi)^2$.
Terms involving field strength tensors, such as $F_{\mu\nu}F^{\mu\nu}$ or $\sigma_{\mu\nu}F^{\mu\nu}$, usually come with a factor of $\kappa^4$ in renormalisable field theories.

\begin{figure}[h]
  \begin{center}
  \includegraphics[width=.25\textwidth]{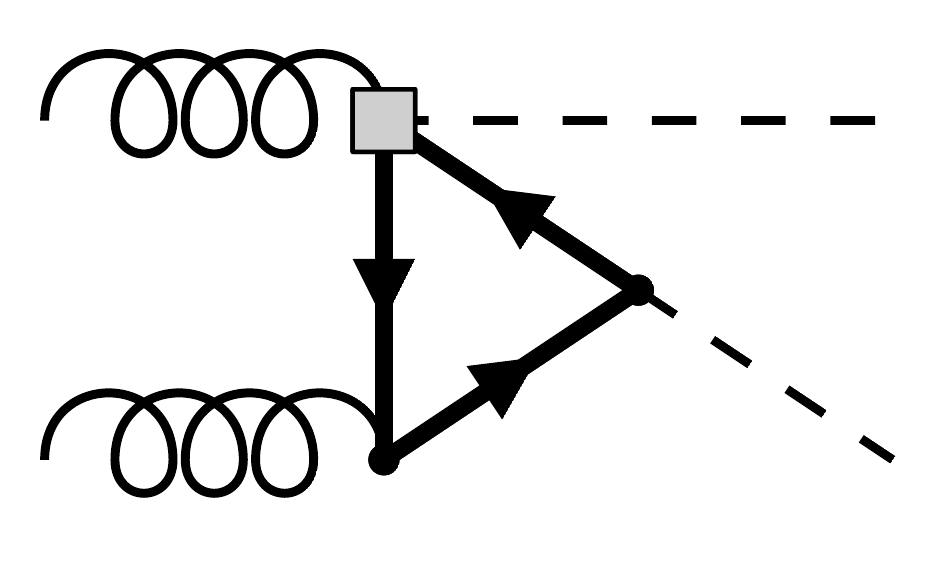}%
  \includegraphics[width=.35\textwidth]{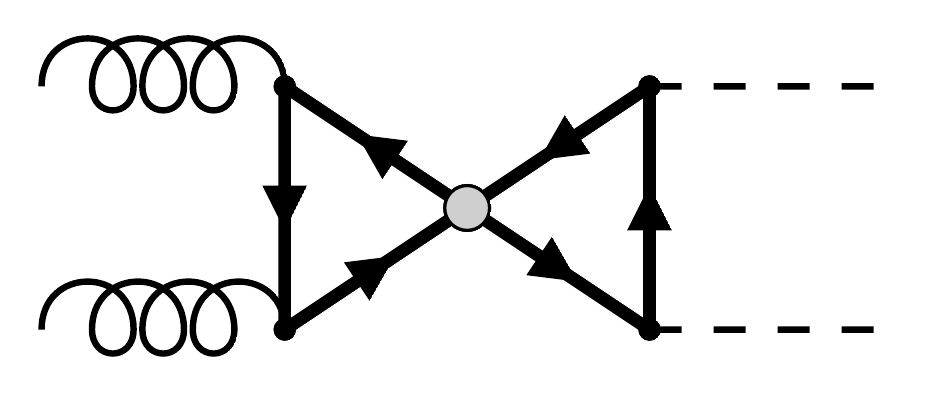}
 \caption{Example diagrams for operators contributing to $gg\to hh$ in SMEFT at order $\frac{1}{\Lambda^2}\left( \frac{1}{16\pi^2}\right)^2$. \label{fig:chromo_4f_diagrams}}
 \end{center}
 \end{figure}

Thus we find $d_\chi=6$ for the chromomagnetic operator ${\cal O}_{tG}$.
Applying these results to the process $gg\to hh$, where  ${\cal O}_{tG}$ is inserted into a one-loop diagram, this contribution comes with a factor $\frac{1}{\Lambda^2}\left( \frac{1}{16\pi^2}\right)^2$.
The four-fermion operators are generically tree-generated operators, however in the $gg\to hh$ process they enter for the first time at two-loop order and therefore contain explicit loop suppression factors. This means that diagrams with ${\cal O}_{tG}$ and diagrams with four-top operators enter $gg\to hh$ at the same order, which is $\frac{1}{\Lambda^2}\left( \frac{1}{16\pi^2}\right)^2$.
This power counting  is corroborated by the fact that the connection between these operators becomes crucial to achieve a $\gamma_5$-scheme-independent result~\cite{DiNoi:2023ygk}.
Representative diagrams are shown in Fig.~\ref{fig:chromo_4f_diagrams}.

 The diagram  on the left-hand side contains ${\cal O}_{tG}$, the grey square representing a loop-generated operator inserted into a one-loop diagram,
 while diagram on the right contains a four-fermion operator, the grey circle denoting a tree-generated operator inserted into a two-loop diagram.

 \section{Results}

 In this section we show how the inclusion of the chromomagnetic operator impacts the predictions for the invariant mass of the Higgs boson pair, $\mhh$. Results including four-fermion operators will be shown elsewhere.
 The results have been produced for a center-of-mass energy of $\sqrt{s}=13.6$\,TeV with
 the PDF4LHC15{\tt\_}nlo{\tt\_}30{\tt\_}pdfas~\cite{Butterworth:2015oua}
 parton distribution functions, interfaced via LHAPDF~\cite{Buckley:2014ana}.
 The masses of the Higgs boson and the top quark have been fixed to $m_h=125$\,GeV, $m_t=173$\,GeV and their widths
have been set to zero.
Jets are clustered with the anti-$k_T$ algorithm~\cite{Cacciari:2008gp} as
implemented in the FastJet package~\cite{Cacciari:2011ma}, with jet radius $R=0.4$ and a minimum transverse momentum
$p_{T,\mathrm{min}}^{\rm{jet}}=20$\,GeV. For the central renormalisation and factorisation
scale we use $\mu_R=\mu_F=m_{hh}/2$. 
 
\begin{figure}[h]
   \begin{center}
     \begin{subfigure}[]{0.48\textwidth}
       \includegraphics[width=\textwidth,page=1]{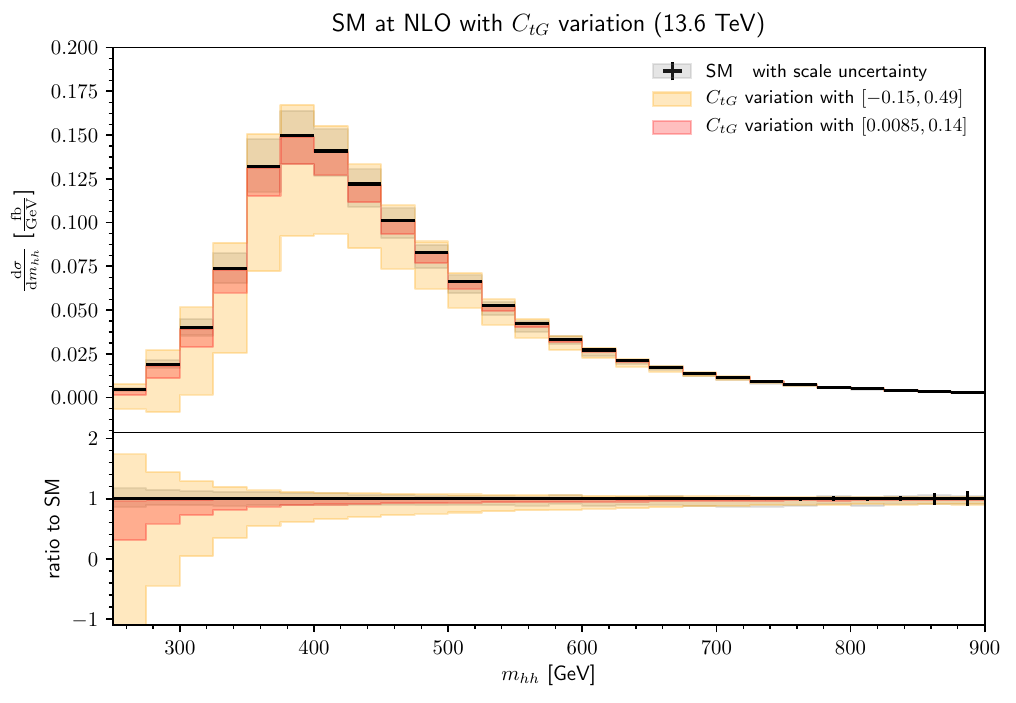}%
       \subcaption{\label{fig:nlo_CtG} Higgs boson pair invariant mass distribution at $\sqrt{s}=13.6$\,TeV with variations of the Wilson coefficient for the chromomagnetic dipole operator, ${\cal O}_{tG}$. The ranges of  are taken from Ref.~\cite{Ethier:2021bye} based on an ${\cal O}(\Lambda^{-2})$ fit, where the range for the yellow curve stems from a fit marginalised over the other Wilson coefficients, the range for the red curve from an individual fit.}   \end{subfigure} \hfill
     \vspace*{-5mm}
       \begin{subfigure}[]{0.49\textwidth}
       \includegraphics[width=\textwidth,page=1]{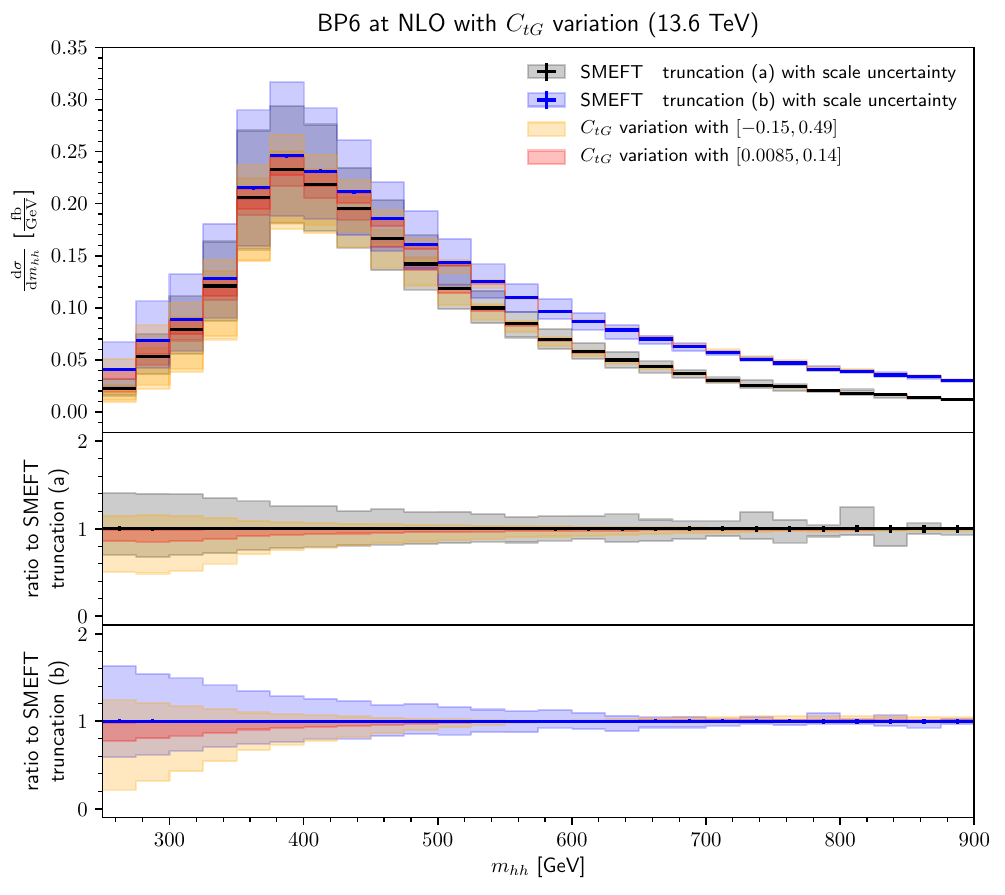}%
     \subcaption{\label{fig:nlo_bp6_test_CtG} Benchmark point 6$^\star$ with and without inclusion of ${\cal O}_{tG}$, for (a) linear and (b) quadratic truncation options.}
    \end{subfigure} 
\end{center}
\end{figure}

Fig.~\ref{fig:nlo_CtG} shows the effect of the chromomagnetic operator
in isolation, all other couplings have been set to the SM values. The
two variation ranges for $C_{tG}$ are inspired by
Ref.~\cite{Ethier:2021bye}, where constraints have been extracted from
global fits of Higgs, diboson and top quark observables. We use two ranges obtained from a ${\cal O}(\Lambda^{-2})$ fit: the smaller range is based on an individual fit of  $C_{tG}$ with all other parameters set to SM values, the larger range is the result of a fit marginalised over the other Wilson coefficients.
For the $\mhh$ distribution, variations in both ranges are larger than the SM scale uncertainties at low $\mhh$ values, for the smaller variation range (red curve) only for $\mhh$ values below the top quark pair production threshold, leading to a decrease of the differential cross section by about 60\% in the bin at the Higgs boson pair production threshold.

Fig.~\ref{fig:nlo_bp6_test_CtG} shows the $\mhh$ distribution for benchmark point 6$^\star$, defined by
$\chhh=-0.684, \ct=0.9, \ctt=-1/6, \cg=0.5, \cgg=0.25$ with and without inclusion of the chromomagnetic operator.
Truncation option (a) denotes the truncation of the expansion in
$1/\Lambda$ at order $1/\Lambda^2$ (dimension-6) at cross section
level, truncation option (b) includes the squared dimension-6 terms at
cross section level~\cite{Heinrich:2022idm}. Note that the chromomagnetic operator itself will only enter linearly in both, truncation options (a) and (b), because its square would be of higher order in the loop counting.

For benchmark point 6$^\star$, the inclusion of $C_{tG}$, varied in the range given by the individual fit (red band), also leads to  a decrease of the differential cross section at low $\mhh$ values.  Such an effect could be compensated by values of $\chhh$ larger than one, as the latter tend to enhance the low-$\mhh$ region. Such a combination could produce $\mhh$ distributions where the shape is degenerate to the SM shape.

\section{Conclusions}

We have discussed the power counting in SMEFT and HEFT and analysed the loop suppression factors of operators contributing to Higgs boson pair production in gluon fusion. Under the assumption of a renormalisable, weakly coupled theory beyond the scale $\Lambda$, the chromomagnetic operator as well as four-fermion operators have been identified to enter this process at the same order in the EFT counting scheme.
The impact of the chromomagnetic operator on the $\mhh$ distribution, with its coefficient varied in a range motivated by current LHC constraints, is strongest in the low-$\mhh$ range and  decreases the differential cross section if all other parameters are kept SM-like.

\section*{Acknowledgements}

We would like to thank Stephen Jones and  Matthias Kerner for
collaboration related to the $ggHH$@NLO code and Gerhard Buchalla  for useful discussions. 
This research was supported by the Deutsche Forschungsgemeinschaft (DFG, German Research Foundation) under grant 396021762 - TRR 257.


\bibliographystyle{JHEP}

\bibliography{radcor2023}


\end{document}